\documentclass[12pt,onecolumn]{IEEEtran}
 
\usepackage[colorlinks,bookmarksopen,bookmarksnumbered,citecolor=red,urlcolor=red,]{hyperref}
 \usepackage{indentfirst}
\usepackage{algorithm,algorithmic,amsbsy,amsmath,amssymb,epsfig,bbm,mathrsfs, bbm}  
\usepackage{multirow}
\usepackage{mathrsfs}
 \usepackage{epstopdf}

\usepackage{verbatim}
\usepackage{amsfonts}
\usepackage{amsthm}
\begin{document}
\pagestyle{empty}
\begin{titlepage}

\title{A Layered Coalitional Game Framework of Wireless Relay Network}

\author{Xiao Lu, Ping Wang, Dusit Niyato \\
   ~School of Computer Engineering, Nanyang Technological University, Singapore\\
  Email: \{Luxiao, Wangping, Dniyato\}@ntu.edu.sg
  }

\markboth{Transactions on Vehicular Technology, accepted for publication}{Shell \MakeLowercase{\textit{et al.}}: Bare Demo of
IEEEtran.cls for Journals}

\maketitle

\begin{abstract}

A wireless relay network (WRN) has recently emerged as an effective way to increase communication capacity and extend a coverage area with a low cost. In the WRN, multiple service providers (SPs) can cooperate to share their resources (e.g., relay nodes and spectrum), to achieve higher utility in terms of revenue. Such cooperation can improve the capacity of the WRN, and thus throughput for terminal devices (TDs). However, this cooperation can be realized only if fair allocation of aggregated utility, which is the sum of the utility of all the cooperative SPs, can be achieved. In this paper, we investigate the WRN consisting of SPs at the upper layer and TDs at the lower layer and present a game theoretic framework to address the cooperation decision making problem in the WRN. Specifically, the cooperation of SPs is modeled as an overlapping coalition formation game, in which SPs should form a stable coalitional structure and obtain a fair share of the aggregated utility. We also study the problem of allocating aggregated utility based on the concept of Shapley value, which stabilizes the cooperation of SPs in the WRN. The cooperation of TDs is modeled as a network formation game, in which TDs establish links among each other to form a stable network structure. Numerical results demonstrate that the proposed distributed algorithm obtains the aggregated utility approximating the optimal solutions and achieves good convergence speed.
\end{abstract}

\emph{Keywords-} Layered coalitional game, overlapping coalition formation, network structure formation, utility allocation solution, Shapley value.

\IEEEpeerreviewmaketitle
\end{titlepage}

\pagestyle{plain} 
\pagenumbering{arabic} 
\setcounter{page}{1}
\section{Introduction}

The multi-hop relay transmission has potential to offer low-cost, high-speed and long-range wireless communications. The multi-hop relay transmission will be adopted in the new standards (e.g., WiMAX or IEEE 802.16m). In such standards, the multi-hop relay capability is enabled for legacy network terminal devices (TDs) (e.g., access points and routers). In this context, the wireless relay network (WRN) has recently emerged as one of the focuses on next generation wireless networks. In the WRN, relay transmission based on cooperation among TDs can mitigate channel fading and exploit spatial reuse to accommodate concurrent transmissions, thus potentially increasing network capacity and coverage. Existing works~\cite{Sadek2007,Boyer2004} have demonstrated that a significant performance improvement can be achieved in the WRNs in terms of higher throughput, lower bit error rate, better capacity, compared to traditional single-hop transmission used in the current cellular network~\cite{16m2007}.

This paper considers the multi-hop network structure formed by TDs from multiple service providers (SPs) with the purpose of capacity improvement. Multiple SPs may form coalitions to let their TDs serve the transmission demand of each other, with the goal to increase the SPs' aggregated revenue~\cite{Lin10}. The utilization of available resources, i.e., vacant TDs (which can relay the data for other TDs) and network capacity, can be substantially improved through forming SP coalitions. This is beneficial to SPs because, in some WRNs, TDs are deployed dispersedly. For each SP, if there is no cooperation, the limited number of available vacant TDs and their dispersive locations largely confine the link capacity for data transmission. The cooperation among SPs could increase the number of available vacant TDs for each SP which in turn can improve network capacity. Accordingly, a larger amount of traffic demand can be served by more efficiently resource utilization in the WRN, which leads to higher aggregated revenue for SPs. The cooperation in the WRN also benefits TDs because a cooperative relay may substantially lead to improved network capacity and thus increased throughput for the TDs. We call the formation of this interrelated SP cooperation and TD cooperation as the layered cooperation problem, which is the main focus of this paper. Previous work has also considered game-theoretic framework with layers/hierarchies, e.g., \cite{Xiao2012} in the cognitive radio networks, \cite{Kang2012} in two-tier femtocell networks and \cite{cao2013} in WRNs. However, most of the works considered the competition relationship between different layers, which belongs to the Stackelberg game concept \cite{Fudenberg1993}. In our work, different layers interact to improve the benefit of each other cooperatively.

Although cooperation in the WRN brings significant benefits, three major challenges arise. First, what is the coalitional structure desirable for all SPs? Second, how shall the aggregated utility (i.e., revenue) be allocated to each SP in the coalition in a fair manner so that none of SPs is willing to leave the coalition? Third, what is the stable network structure for TDs of cooperative SPs to perform uplink transmission? This paper addresses these three challenges.

\newtheorem{definition}{Definition}

\section{Layered Coalitional Game of SPs and TDs in WRN}
We consider a multi-cell WRN consisting of multiple BSs and TDs deployed by different SPs. Let $\mathcal{M}=\{ 1, 2,\ldots,M\}$ and $\mathcal{B}=\{ 1, 2,\ldots,B\}$ denote the sets of SPs and BSs, respectively. $\mathcal{N}^{(m)}=\{ 1, 2,\ldots,N_{m}\}$ is the set of TDs of SP $m$, and $N_{m}$ is the total number of TDs belonging to SP $m$, for $m \in \mathcal{M}$. We call the TD with and without transmission demand at a specific time as the source TD and vacant TD, respectively. In the WRN, multi-hop relaying is used to route the flow from the source TD to the destination BS. We assume that each TD $i$ is connected to a BS through at most one path and each vacant TD can only work as a relay for one TD.

We formulate a layered game theoretic model, referred to as the layered coalitional game, to model the decision-making process of cooperation between the SPs and TDs. The SPs aim to maximize their individual utility, while the TDs attempt to maximize their end-to-end throughput with the help of TDs from cooperative SPs. In the proposed game, SPs need to decide on the coalitional structure to form to improve their individual as well as aggregated utility. Also, the TDs need to make decisions to form a relay network structure to improve throughput. The utility of an SP is defined as the revenue (earned for meeting the transmission demand of TDs) subtracted by TD cost and coalition cost. The TD cost accounts for the energy consumption of TDs in transmitting and forwarding packets. The coalition cost accounts for the factors such as communication cost (i.e., packet overhead for information exchange between TDs from different SPs) and operation cost (i.e., energy consumed for information exchange). Recognizing the behavior of SPs and TDs, we propose to use the overlapping coalition formation game to model the behavior of SPs and the link formation game to characterize the interaction among TDs.

\subsection{Overlapping Coalition Formation Game} \label{sec_ocf}
An overlapping coalition formation game is formulated at the SP layer whose rational players are the SPs in set $\mathcal{M}$. The coalition denoted by $\mathcal{T}$ is a subset of $\mathcal{M}$, and $\mathcal{M}$ is the grand coalition (i.e, coalition of all SPs). The \emph{strategy} of the SP is to form the coalitions to improve the aggregated utility. Note that different from non-overlapping coalition formation game where players have to cooperate with all others in the same coalition and each player only stays in one coalition, in the overlapping coalition formation game, each player is able to cooperate and join multiple different coalitions. Thus, a non-overlapping coalition can be considered as the two-player sub-coalitions. For example, a three-player coalition $\{1,2,3\}$ is equivalent to three sub-coalitions $\{1,2\},\{1,3\},\{2,3\}$ in the overlapping coalition formation game.

The characteristic function $v(\mathcal{T})$ of coalition $\mathcal{T}$ can be defined as follows:
\begin{eqnarray} \label{allocation}
v(\mathcal{T})=\sum_{m \in\mathcal{T}}\phi_{m}(\mathcal{T}),
\end{eqnarray}
and $v(\emptyset)=0$, where $\phi_{m}(\mathcal{T})$ is the allocated utility of SP $m$ in coalition $\mathcal{T}$.

Let $\Gamma_{m}$ denote the set of coalitions that SP $m$ belongs to. The total utility function of an individual SP $m \in \Gamma_{m}$ can be represented as follows:
\begin{eqnarray}
v_{m}(\Gamma_{m})=\sum_{\mathcal{T} \in \Gamma_{m}} \phi_{m}(\mathcal{T}).
\end{eqnarray}

A coalitional structure can be defined in a characteristic function as the set $\omega = \{ \mathcal{T}_{1},\mathcal{T}_{2},\ldots,\mathcal{T}_{l} \}$ which is a collection of non-empty subsets of $\mathcal{M}$ such that $\bigcup^{l}_{k=1} \mathcal{T}_{k} = \mathcal{M}, \forall k, \mathcal{T}_{k} \subseteq \mathcal{M} $ and the sub-coalitions could overlap with each other. $l$ is the total number of coalitions in collection $\omega$. The total number of possible collections for the $M$-SP game is denoted by $D_{M}$ where
\begin{eqnarray} \label{number}
& D_{M} = \left\{
\begin{array}{lcl}
 M , & & \text{$M=1,2$} , \\
(M-1)^{M} , & & \text{otherwise}.
\end{array} \right.
\end{eqnarray}

Let $\Lambda$ be the set of all covers of $\mathcal{M}$. The cover function on $\mathcal{M}$ is a function $v$ such that
\begin{eqnarray}
v: 2^{|\mathcal{M}|}\times \omega \to \mathbb{R}. \hspace{5mm} (\mathcal{T}, \omega) \mapsto v(\mathcal{T}, \omega), \hspace{3mm} \mathcal{T} \in \Lambda. \nonumber
\end{eqnarray}

A coalition structure can be either \emph{stable} or \emph{unstable}. If a coalitional structure is stable, it will remain still. Otherwise, the coalitional structure will transit to any other state.  A coalitional structure $\omega$ is unstable under following cases: $(i)$ if there exists an SP $m \in \Gamma_{m}$ which can get higher utility if it does not make any cooperation with others, i.e., $v(\{m\}) > \phi_{m}(\Gamma_{m})$; $(ii)$, if there exists an SP  $m \in  \mathcal{T} \in \Gamma_{m}$ which can get higher utility by leaving the coalition $\mathcal{T}$, i.e., $\phi_{m}(\Gamma_{m} / \mathcal{T}) \geq \phi_{m}(\Gamma_{m})$;
$(iii)$, if there exists an SP $m \in \Gamma_{m}$, which is not a member of coalition $\mathcal{T}$, can get higher utility by joining coalition $\mathcal{T}$, i.e., $\phi_{m}(\mathcal{T} \cup \Gamma_{m}) > \phi_{m}(\Gamma_{m})$.

The solution of the proposed overlapping coalition formation game is the stable coalitional structure for SPs. A given coalitional structure $\omega$ is stable if it shows both the \emph{internal stability} as well as \emph{external stability}~\cite{Arnold2002}. We define the concepts of internal stability and external stability of stable overlapping coalition as follows:

\begin{itemize}
\item \textbf{Internal stability}: If coalition $\mathcal{T} \subseteq \mathcal{M}$ is internally stable, none of the SPs can improve its utility by leaving this coalition $\mathcal{T}$ or by joining another coalition $\mathcal{U} \subseteq \mathcal{M} $ and $\mathcal{U} \neq \mathcal{T}, $ i.e., $\phi_{m}(\Gamma_{m}) \geq v(\{m\})$, $\phi_{m}(\Gamma_{m}) \geq \phi_{m}(\Gamma_{m} \setminus \mathcal{T} )$
, $\forall m \in \mathcal{T}, \mathcal{T} \in \Gamma_{m}$ and $ \mathcal{U} \notin \Gamma_{m}$.

\item \textbf{External stability}: If coalition $\mathcal{T}$ is externally stable, there is no SP $m$ from other coalition set $\Gamma_{m}$ that can improve its utility by joining a coalition $\mathcal{T}$, i.e., $\phi_{m}(\Gamma_{m}) > \phi_{m}(\mathcal{T} \cup \Gamma_{m})$ for $m \in \mathcal{T} \subseteq \mathcal{M}$, $ \mathcal{T} \notin \Gamma_{m}$.

\end{itemize}

The internal stability prevents to get unstable coalitional structures like cases ($i$) and ($ii$), while the external stability prevents the happening of case ($iii$). Based on these concepts of stability,
we propose a distributed overlapping coalition formation algorithm consisting of ``merge" and ``split" methods that are able to update the coalitional structure $\omega$ of the set $\mathcal{M}$.
Let $\phi(\omega)$ and $\phi_{m}(\omega)$ denote the aggregated utility of all the SPs under coalitional structure $\omega$ and the allocated utility of the SP $m$ associated with the coalitional structure $\omega$, respectively. Given a current set of coalitions, any SP can join or split from any coalition to increase their individual utility. Thus, the SP $m$ belonging to a coalition set $\Gamma_{m}$ may also join coalition $\mathcal{T} \notin \Gamma_{m}$ if this action leads to a strict increase in its allocated utility, while at the same time, does not lead to any decrease in the utility of SPs already in coalition $\mathcal{T}$. Mathematically, we have,

\textbf{Merge rule}: The SP $m \in \Gamma_{m}$ can make cooperation with another SP $n \notin \Gamma_{m}$ to form a new coalition $\mathcal{T}$ if the following conditions are satisfied,
\begin{eqnarray}
\phi_{m}( \Gamma_{m}\cup \mathcal{T}) > \phi_{m}(\Gamma_{m}), \hspace{3mm} m \in \mathcal{T}, \hspace{3mm} m \in \Gamma_{m}, \nonumber \\
\text{and} \hspace{3mm} \phi_{n}(\Gamma_{n} \cup \mathcal{T}) \geq \phi_{n}(\Gamma_{n}), \hspace{3mm} n \in \mathcal{T}, \hspace{3mm} n \notin \Gamma_{m}.
\end{eqnarray}

In general, with the coalitional structure $\omega_{k}$, the SP $m \in \Gamma_{m}$ can cooperate with other SPs to form a new coalition $\mathcal{T}$, changing its coalition set to $\Gamma^{\prime}_{m}$ with the new coalitional structure represented as $\omega^{\ddag}_{k^{\prime}}$, if the following condition is satisfied,
\begin{eqnarray}
\phi_{m}(\omega^{\ddag}_{k^{\prime}}) > \phi_{m}(\omega_{k}), \hspace{3mm} \forall m \in \mathcal{T}, \hspace{3mm} \Gamma_{m} \subseteq \omega_{k}, \hspace{3mm} \Gamma^{\prime}_{m} \subseteq \omega^{\ddag}_{k^{\prime}}. \nonumber
\end{eqnarray}
In particular, multiple SPs from multiple sets of coalitions will merge and become a new coalition if the SPs in the formed coalition receive higher utility.

Also, the SP $m \in \Gamma_{m}$ would break cooperation with another SP $n \in \Gamma_{m}$ if this action leads to a strictly better individual utility. Thus, we have,
\textbf{Split rule}: The SP $m \in \mathcal{T} \in \Gamma_{m}$ would break cooperation with another SP $n \in \mathcal{T}$
if the following condition is satisfied,
\begin{eqnarray}
\phi_{m}(\Gamma_{m} \setminus \mathcal{T} ) > \phi_{m}(\Gamma_{m}), \hspace{3mm} \mathcal{T} \in \Gamma_{m}, \hspace{3mm} m \in \Gamma_{m}. \nonumber
\end{eqnarray}

In general, the coalitional structure $\omega_{k}$, SP $m \in \Gamma_{m}$ can break cooperation with other SPs from coalition $\mathcal{T} \in \Gamma_{m}$, changing its coalition set to $\Gamma_{m}^{\prime\prime}$ with the new coalitional structure represented as $\omega^{\dag}_{k}$
if the following condition is satisfied,
\begin{eqnarray}
\phi_{m}(\omega^{\dag}_{k^{\prime}}) > \phi_{m}(\omega_{k}), \hspace{3mm} \exists m \in \mathcal{T}, \hspace{3mm} \Gamma_{m} \subseteq \omega_{k}, \hspace{3mm} \Gamma^{\prime\prime}_{m} \subseteq \omega^{\dag}_{k^{\prime}} .
\end{eqnarray}
In particular, the SP in a coalition agrees to split from the coalition if the utility of the SP is higher than that in the coalition.

Denote the coalitional structure at time $t$ by $\omega(t)$.
Given the current coalitional structure at time $t$, a new coalitional structure of SP $m$ can be updated following an updating rule based on the merge-and-split algorithm to maximize the instant utility. We define the update process as follows:
\begin{eqnarray}
& \omega_{k}(t+1)= \left\{
\begin{array}{rcl}
\omega^{\dag}_{k^{\prime}} , & & { \phi_{m}(\omega^{\dag}_{k^{\prime}})> \phi_{m}(\omega^{\ddag}_{k^{\prime}}), \hspace{3mm} \text{and} \hspace{3mm} \phi_{m}(\omega^{\dag}_{k^{\prime}})> \phi_{m}(\omega_{k}(t))}, \\
\omega^{\ddag}_{k^{\prime}},& &{\phi_{m}(\omega^{\ddag}_{k^{\prime}})> \phi_{m}(\omega^{\dag}_{k^{\prime}}), \hspace{3mm} \text{and} \hspace{3mm} \phi_{m}(\omega^{\ddag}_{k^{\prime}})> \phi_{m}(\omega_{k}(t))}, \\
\omega_{k}(t) , & & \text{otherwise}.
\end{array} \right.
\end{eqnarray}

With this merge-and-split updating rule, we propose the overlapping coalition formation algorithm for SPs in \textbf{Algorithm 1}. For dynamically changing environment (e.g., TD failure, topology change or traffic demand change), the link formation algorithm, i.e., \textbf{Algorithm 2}, is run periodically to update link information to SPs. This period between each time that \textbf{Algorithm 2} is run will be based on the network environment dynamics. The more dynamic the network environment is, the smaller the period should be, so that the network structure can adapt to the environment change quickly. Then, the SPs accordingly adapt to environment changes by adjusting a coalitional structure.

\begin{algorithm}[width=\textwidth]
\caption{Overlapping Coalition Formation Algorithm}
\label{CFA}
{\fontsize {9}{9}\selectfont
\begin{algorithmic}
\STATE \textbf{Initial State} \\
\STATE \hspace{3mm} In the starting network, each SP $m$ either has no cooperation with others or serves as a member in a  coalition set $\Gamma_{m}$.
\STATE \textbf{Coalition Formation Process} \\
\STATE \hspace{3mm} \textbf{Phase 1} \emph{Discovery} \\
\STATE \hspace{6mm} According to the \textbf{Initial State}, TDs from each SP perform \emph{Dynamic Virtual Link Replacement} algorithm specified in \textbf{Algorithm 2}.
\STATE \hspace{6mm} Based on the information feedback from TD layer, each SP $m$ calculates the corresponding aggregated utility of each coalition
\STATE \hspace{6mm} $\mathcal{T} \in \Gamma_{m}$, and its share $\phi_{m}(\Gamma_{m})$, $m\in \Gamma_{m}$  based on some utility allocation method.
\STATE \hspace{3mm} \textbf{Phase 2} \emph{Overlapping Coalition Formation} \\
\STATE \hspace{6mm} SPs perform overlapping coalition formation (i.e., either cooperate or break cooperation) using merge-and-split algorithm. During
\STATE \hspace{6mm} this process, SPs which attempt to make cooperation let their TDs perform the \emph{Dynamic Virtual Link Replacement} algorithm and
\STATE \hspace{6mm}  provide feedback to their SPs. Based on the information feedback, each SP calculates its potential share  $m\in \Gamma_{m}$ $\phi^{\prime}_{m}(\Gamma_{m})$,
\STATE \hspace{6mm} when cooperation is reached and the SP compares it with the SP's current utility.
\STATE \hspace{9mm} \textbf{repeat}
\STATE \hspace{12mm} (a) $\omega_{k}=\omega^{\ddag}_{k^{\prime}}$: SPs decide to make cooperation based on the \textbf{Merge Rule}.
\STATE \hspace{12mm} (b) $\omega_{k}=\omega^{\dag}_{k^{\prime}}$: SPs decide to break cooperation based on the \textbf{Split Rule}.
\STATE \hspace{9mm} \textbf{until} merge-and-split algorithm converges.
\STATE \hspace{3mm} \textbf{Phase 3} \emph{Cooperative Sharing} \\
\STATE \hspace{6mm} The SPs in the same coalition share their vacant TDs with each other for link-layer transmission.
\end{algorithmic}}
\end{algorithm}

The Markov chain defined by the state space of all the coalitional structures can be used to represent the overlapping coalition formation process.
The state space
can be denoted as $\bigtriangleup=\{(\omega_{x});x=\{1,\ldots, D_{M}\}\}$, where $\omega_{x}$ represents the coalitional structure of all SPs and $D_{M}$ can be obtained from (\ref{number}). Once all SPs reach the state in the set of absorbing states in the Markov chain, they will remain in the set of absorbing states forever. The transition from one state to another state takes place based on the merge-and-split decision taken by the SPs. $\omega \triangleright \omega^{\prime}$ is the reachability condition. In particular, if collection $\omega^{\prime}$ is reachable from $\omega$ given the decision of all SPs defined in Section V, the condition $\omega \triangleright \omega^{\prime}$ is true and false otherwise. In addition, once the absorbing state is reached, the state transition will stop. This absorbing state is the stable coalitional structure $\omega^{\star}$, given that none of the SPs has an incentive to change its decision. Such a stable coalitional structure is the solution to our overlapping coalition formation game.

\subsection{Link Formation Game} \label{sec_nsf}

The next step in the proposed layered coalitional game is to find a link formation algorithm that can model the interactions among the TDs seeking to form a network structure to maximize the payoff (i.e., the end-to-end throughput of all the supporting flows).

The \emph{action space} for each transmitting TD $i$ includes the vacant TDs or BSs that the transmitting TD $i$ can connect to as the next hop. The action space for each receiving TD consists of TDs which this receiving TD can establish a link from. When the receiving TD accepts a relay request from another TD, the receiving TD then becomes the transmitting TD and plays actions as a transmitting TD does, i.e., trying to find the next hop to improve its payoff. Note that a TD $i$ cannot connect to another TD $j$ which is already connected to $i$, i.e., if $(j,i) \in G$, then $(i,j) \notin G$. Note that if a source TD fails to find a vacant TD to form a link, this TD will directly connect to a BS for uplink transmission. In this regard, prior to performing the link formation algorithm, initially all the source TDs have direct link connections to the BSs.

Please note that although a transmitting TD $i$ establishes a link to a receiving TD $j$ from its action space, the payoff of TD $i$ may not always be improved. Therefore, the rational action of each transmitting TD $i$ is to select a link $(i,j)$, that this transmitting TD $i$ wants to establish from its action space, such that its payoff is improved. For the receiving TD $j$, the strategy is to choose a link $(i,j)$, that this receiving TD $j$ wants to accept from its action space to improve its payoff.

We assume that all the TDs play their actions one by one. That is, when a TD $i$ plays an action, all the other TDs do not take new actions. Since only one TD can be connected to as a relay, forming a new link $(i,j)$ for a relay TD $j$ also implies disconnecting its previous link (if any). That is, if TD $i$ allows TD $j$ to act as a relay when link $(k,j)$ already exists, the TD $j$ will abandon link $(k,j)$ and establish the new link $(i,j)$ with the transmitting TD $i$ as long as the condition $ v(i \cup j)- v(i) > v(k \cup j) - v(k) $ is satisfied. This is referred to as \emph{link replacement}.

Although each TD $i \in \mathcal{N}$ can play any action, there may exist some link $(i,j)$ that cannot be established. For example, a vacant TD $j$ may not accept the formation of $s_{i}$, if it has another better option, i.e., there exists a link $(k,j)$ satisfying the condition $v(k \cup j)- v(k) > v(i \cup j) - v(i)$. Hence, a \emph{feasible strategy} for a transmitting TD $i$ is a link $(i,j)$ where the receiving TD $j$ is willing to accept from the TD $i$, i.e., $v(i \cup j)-v(i) > 0$ and $\nexists (k,j)$ satisfies $v(k \cup j)> v(i\cup j)$. Also, the feasible strategy for a vacant TD $j$ is a link $(i,j)$ that the transmitting TD $i$ wants to establish with TD $j$, i.e., $v(i \cup j)-v(i) > 0$ and $\nexists (i,k)$ satisfies $v(i \cup k)> v(i\cup j)$. We denote $\mathcal{S}_{i}$ the feasible strategy space which consists of all the strategies of TD $i$. When the TD $i$ plays strategy $s_{i}$, while all other TDs maintain their current strategies denoted by a vector $\textbf{s}_{-i}$, the resulting network graph is denoted by $G_{s_{i},\textbf{s}_{-i}}$.

We further consider that all the TDs are myopic in the sense that each TD responds to improve the payoff given the current strategies of the other TDs, without the foresight of the future evolution of the network. Based on this, we define the concept of \emph{best response} for TDs as following.

\begin{definition}
A feasible strategy $s^{\star}_{i}\in \widetilde{\mathcal{S}_{i}}$ is a best response for a TD $i \in \mathcal{N}$ if $v_{\{i\}}(G_{s^{\star}, \textbf{s}_{-i}}) \geq v_{\{i\}}(G_{s_{i},\textbf{s}_{-i}})$, $\forall s_{i} \in \widetilde{\mathcal{S}_{i}}$.
\end{definition}

Given a network graph $G$, the best response for a transmitting TD $i$ is to form a link with a vacant TD or BS among all the feasible strategies that maximizes its payoff. On the other hand, the best response for a vacant TD $j$ is to form the link with the transmitting TD $i$ that the TD $j$ can help to achieve the highest increment of throughput, i.e., $\max ( v(i\cup j)-v(i) )$. Because the vacant TD is myopic, it has the belief that the greater payoff it can help to improve, the more utility its SP will share.

Based on the concept of best response, we propose a link formation algorithm shown in \textbf{Algorithm 2}. The link formation algorithm is run periodically, allowing the TDs to adapt to the dynamic environment change.
We define an iteration as a round of sequential plays during which all the TDs have played their myopical strategies to respond to each other. The proposed algorithm can consist of one or more iterations. During each iteration, each TD $i$ chooses to play its current best response $s^{\star}_{i} \in \widetilde{\mathcal{S}_{i}}$ sequentially to maximize its payoff given the current strategies of other TDs. Therefore, this phase is based on the iterative best response of TDs. This feasible best response is indeed an iterative replacement process, as the best strategy of each TD may change over time. When the algorithm converges, it results in a network in which none of the TDs can improve its throughput. This is referred to as a Nash network, which gives the stability concept of the final network graph $G^{\dag}$. Specifically, the Nash network is defined as follows.

\begin{definition} \label{nash}
A network graph $G(\mathcal{N},\mathcal{E})$ with $\mathcal{N}$ denoting the set of all vertices, i.e. TDs and BSs, and $\mathcal{E}$ denoting the set of all edges, i.e, links between pairs of vertices, is the Nash network in its feasible strategy space $\widetilde{\mathcal{S}}_{i}$, $\forall i \in\mathcal{N}$, if no TD $i \in \mathcal{N}$ can improve its payoff by a unilateral change in its strategy $s_{i} \in \mathcal{S}_{i}$.
\end{definition}

In the Nash network, all the TDs choose links based on their best responses and are thus in the Nash equilibrium. Therefore, none of the TDs can improve its payoff by unilaterally changing it current strategy.

\textbf{Theorem 1:} Given any initial network graph $G_{0}$, the dynamic virtual link replacement algorithm of the link formation game will converge to a final network structure after finite number of iterations. The converged network structure is the Nash network in the space of feasible strategies $\widetilde{\mathcal{S}_{i}}$, $\forall i \in \mathcal{N}$.

\begin{algorithm}[width=\textwidth]
\caption{Link Formation Algorithm}
\label{LFG}
{\fontsize {9}{9}\selectfont
\begin{algorithmic}
\STATE \textbf{Initial State} \\
\STATE \hspace{3mm} The starting network is a star network in which each source TD is connected directly to the nearest BS.
\STATE \textbf{Uplink Network Structure Formation Process} \\
\STATE \hspace{3mm} \textbf{repeat}
\STATE \hspace{6mm} \textbf{Phase 1} \emph{Dynamic Virtual Link Replacement}
\STATE \hspace{9mm} The TDs from SPs that attempts to make cooperation play their strategies sequentially in a random order.
\STATE \hspace{12mm} \textbf{repeat}
\STATE \hspace{15mm} 1) During each iteration $y$, every TD $i$ plays its best response $s^{\star}_{i}$,
\STATE \hspace{15mm} 2) The best response strategy of a transmitting TD is to send a link establishment proposal to a relay TD which is able to
\STATE \hspace{15mm} maximize its payoff. The best response strategy of a relay TD is to respond to the proposal, either acceptance or rejection.
\STATE \hspace{15mm} A link replacement operation is  executed if the relay TD is involved in a link. That is, the relay TD splits from its
\STATE \hspace{15mm} current connected TD and replaces it with a new TD that maximizes its payoff.
\STATE \hspace{12mm} \textbf{until} converge to the Nash network structure $G^{\star}_{F}$.
\STATE \hspace{6mm} \textbf{Phase 2} \emph{Feedback}
\STATE \hspace{9mm} Each TD $i$ sends the information about the link $(i,j) \in G^{\star}_{F}$ back to its SP for a coalition formation decision.
\STATE \hspace{3mm} \textbf{until} Acceptance of the convergent network graph by all the SPs which perform \textbf{Algorithm 1}, i.e., merge-and-split operation of
\STATE \hspace{10mm} \textbf{Algorithm 1} converges.
\STATE \hspace{6mm} \textbf{Phase 3} \emph{Multi-hop Transmission}
\STATE \hspace{9mm} Each source TD transmits to a BS with the possible help of relay TDs from its SP and cooperative SPs according to the final
\STATE \hspace{9mm} network structure $G^{\star}_{F}$.
\end{algorithmic}}
\end{algorithm}

\begin{figure}
\centering
\includegraphics[width=1.0\textwidth]{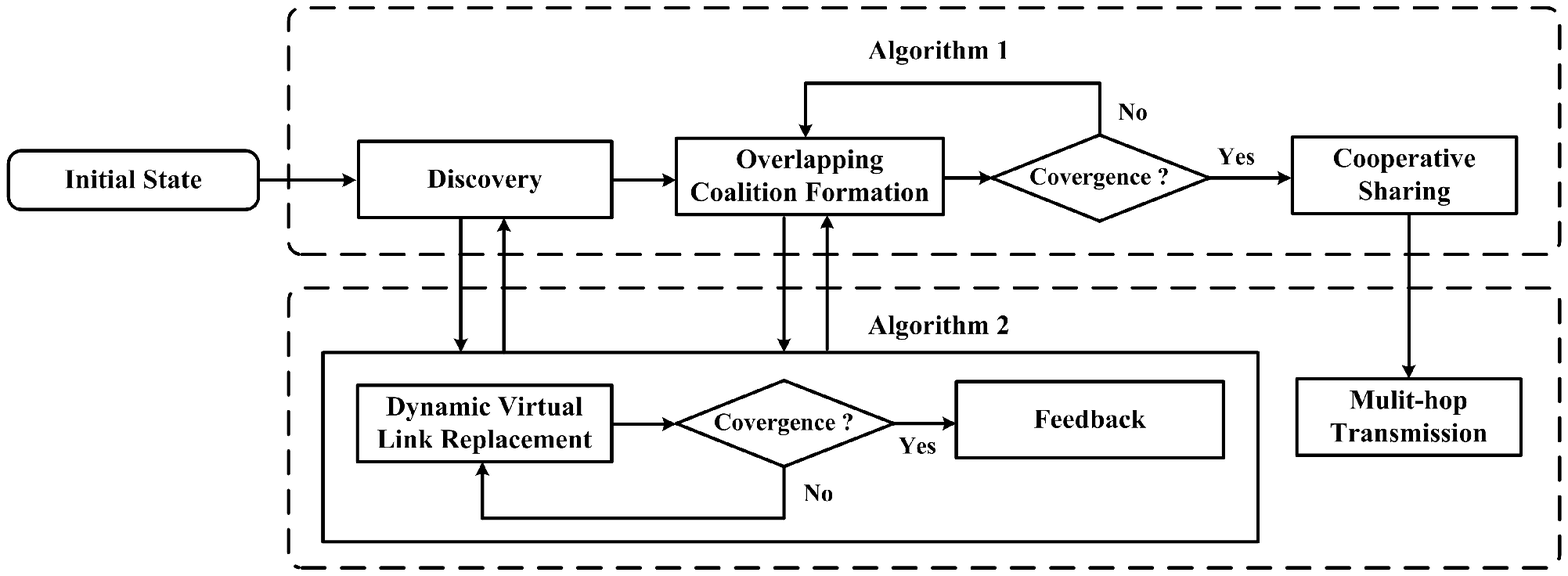}
\caption{Flowchart of the layered coalitional game.} \label{roadmap}
\end{figure}

Due to the space limit, the proof of Theorem 1 is omitted. Figure~\ref{roadmap} shows a flowchart of the proposed layered coalitional game which interrelates \textbf{Algorithm 1} and \textbf{Algorithm 2}. The layered coalitional game starts from the \textbf{Initial State} and firstly goes to the \emph{Discovery} phase of \textbf{Algorithm 1} during which the SPs inform their TD their coalition sets. Then, the algorithm calls the \emph{Dynamic Virtual Link Replacement} and \emph{Feedback} phases of \textbf{Algorithm 2} which provide a convergent network structure according to the initial coalitional structure of SPs. After this, one SP begins to perform overlapping coalition formation with another SP. Also, the SP pair attempting to make cooperation informs their TDs to execute the dynamic virtual link replacement and the TDs provide feedback to the SP pair subsequently. This process is repeated until the merge-and-split operation converges. Finally, the algorithm goes to \emph{Cooperative Sharing} phase where the SPs share their vacant TDs with cooperative SPs and all the TDs perform multi-hop transmission according to the final convergent network structure.

The proposed layered coalitional game can be implemented in a distributed way for WRNs. The decision of each SP can be made by an SP controller (a TD can also be delegated as SP controller in practice), while each TD is responsible for its best response action. For the coalition formation of SPs, each SP controller can communicate with its counterparts and its TDs through wired or wireless communication depending on the network infrastructure. For link formation, each vacant TD can broadcast its information to its neighboring TDs. As the link formation algorithm is run periodically, the period between each algorithm run should be determined considering the  network environment dynamics on a case-by-case basis.

\section{Utility Allocation Solution}

In the proposed layered coalitional game, any reasonable bases for allocating the aggregated utility need to stabilize the overlapping coalitions. In other words, the allocated utility should be imperative to motivate the SPs to join the coalitions which maximize the aggregated utility.
Any set of allocated utility to the SPs are \emph{stable} if it satisfies the following two conditions, i.e.,
\begin{itemize}

\item The sum of the allocated utility for all SPs is equal to the maximum aggregated payoff, i.e., $\sum_{m \in \mathcal{T}} \phi_{m}(\mathcal{T}) = v(\{\mathcal{T}\})$.

\item The allocated utility of each SP should be larger than or equal to the value of the individual SP, i.e., $\phi_{m}(\Gamma_{m}) \geq v(\{m\})$,  $\forall \bigcup_{m \in \Gamma_{m} } m \subseteq \mathcal{M}$.

\end{itemize}

We address the fair utility allocation problem in the SP coalitional game by adopting a well-known concept from static coalitional game theory, i.e., the Shapley value solution. The Shapley value, first introduced in~\cite{Shapley53}, is a unique value based on the marginal contribution of each player to the coalition.

\begin{definition}
The marginal contribution of SP $m$ to a set
$\mathcal{U} \subseteq \mathcal{T} \setminus \{m\}$ is defined as follows:
\begin{eqnarray}
\triangle_{m} (v(\cdot), \mathcal{U}) = v(\mathcal{U} \cup \{m\})-v(\mathcal{U}).
\label{mc}
\end{eqnarray}
The Shapley value for SP $m$ in coalition $\mathcal{T}$ can be obtained from
\begin{eqnarray}
\phi_{m}(\mathcal{T})=\sum_{\mathcal{U} \subseteq \mathcal{T} \setminus \{ m \}}\frac{|\mathcal{U}|!(|\mathcal{T}|-|\mathcal{U}|-1)!}{|\mathcal{T}|!} \triangle_{m}(v(\cdot), \mathcal{U}) .
\label{sv}
\end{eqnarray}
\end{definition}

The Shapley value is suitable for the utility allocation in SP coalitional game due to the properties of efficiency, individual fairness and uniqueness. The efficiency means that the sum of the allocated utility for all SPs equals the maximum aggregated utility, i.e., $\sum_{m \in \mathcal{T}}\phi_{m}(\mathcal{T})=v\{\mathcal{T}\}$. For individual fairness, Shapley value guarantees the allocated utility of each SP to be larger than or equal to the value of the individual SP, i.e., $\phi_{m}(\mathcal{T}) \geq v(\{m\})$, for all $m \in \mathcal{T}$. Uniqueness means that there is only a single result derived by Shapley value solution. The other properties of Shapley value, i.e., symmetry and dummy, can be found in~\cite{winter02}.

In the SP coalitional game, the Shapley value of each SP in a coalition $m$, denoted by $\phi_{m}(\mathcal{T})$, can be derived from the combination of (\ref{mc}) and (\ref{sv}). Then, the total allocated utility of each SP can be obtain from $\phi_{m}(\Gamma_{m})=\sum_{\mathcal{T}\in \Gamma_{m}}\phi_{m}(\mathcal{T})$.

\section{Numerical Results}
In this section, we present the numerical results for the proposed layered coalitional game model. Also, the fairness and stability of the solutions are demonstrated in numerical simulations.

\subsection{Simulation Setting}
For simulation, we consider the uplink transmission of the WRN which adopts TDD-OFDMA scheme. The network consists of two BSs, three SPs, i.e., $\mathcal{M}=\{$O$1$, O$2$, O$3\}$, a number of TDs randomly deployed in a $1km \times 2km$ area. The two BSs locate at the coordinate $(0.5, 0.5)km$ and $(1.5,0.5)km$, respectively. The transmit power of all the TDs is assumed to be the same and fixed at $Q_{i}=10mW, \forall i \in \mathcal{N}$. The noise level is $-90dBm$. We assume that each packet contains $100$ bits of information and no overhead (i.e., $L = M = 100$). The target SINR $\tau_{ij}, \forall i,j \in \mathcal{N}$ is set to $10dB$. For the wireless propagation, we set the path loss exponent $n=4$ and antenna related constant $\beta=62.5$.

For an SP, the revenue gained in a time unit for successfully transmitting a unit normalized throughput (i.e., 1 Kbps) is $120$. The cost for each TD to transmit or forward flow in a time unit is $500$ \emph{per} Watt. We assume that when the SP makes cooperation with another, the fixed cost $C^{co}$ is incurred to both of them. Thus, the total coalition cost afforded by SP $m$ in coalition $\mathcal{T}$ is $\xi^{\mathcal{T}}_{m}(\mathcal{T})= C^{co} (|\mathcal{T}| - 1 )$. The total coalition cost for SP $m$ can be calculated by $\xi_{m}=\sum_{\mathcal{T}\in \Gamma_{m}}\xi^{\mathcal{T}}_{m}$. We set $C^{co}=5$ unless otherwise stated. All the network profiles, i.e., location of source TDs and vacant TDs, are randomly generated.

\subsection{Three-SP Coalition with Uniformly Distributed TDs}

\begin{figure}
\centering
\includegraphics[width=0.8\textwidth]{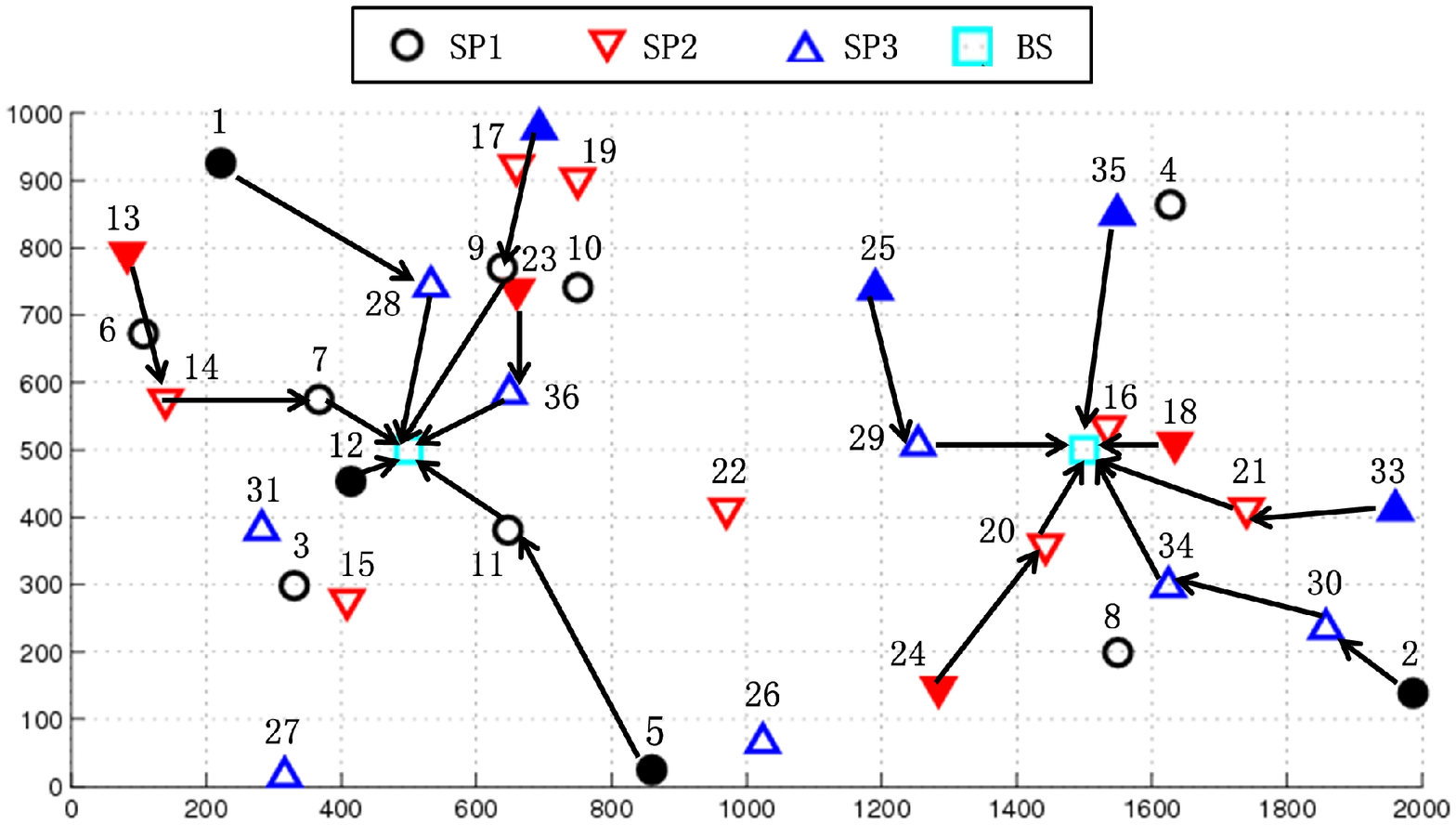}
\caption{Network structure of the WRN (12 Source TDs)} \label{12sources}
\end{figure}

\begin{figure}
\centering
\includegraphics[width=0.8\textwidth]{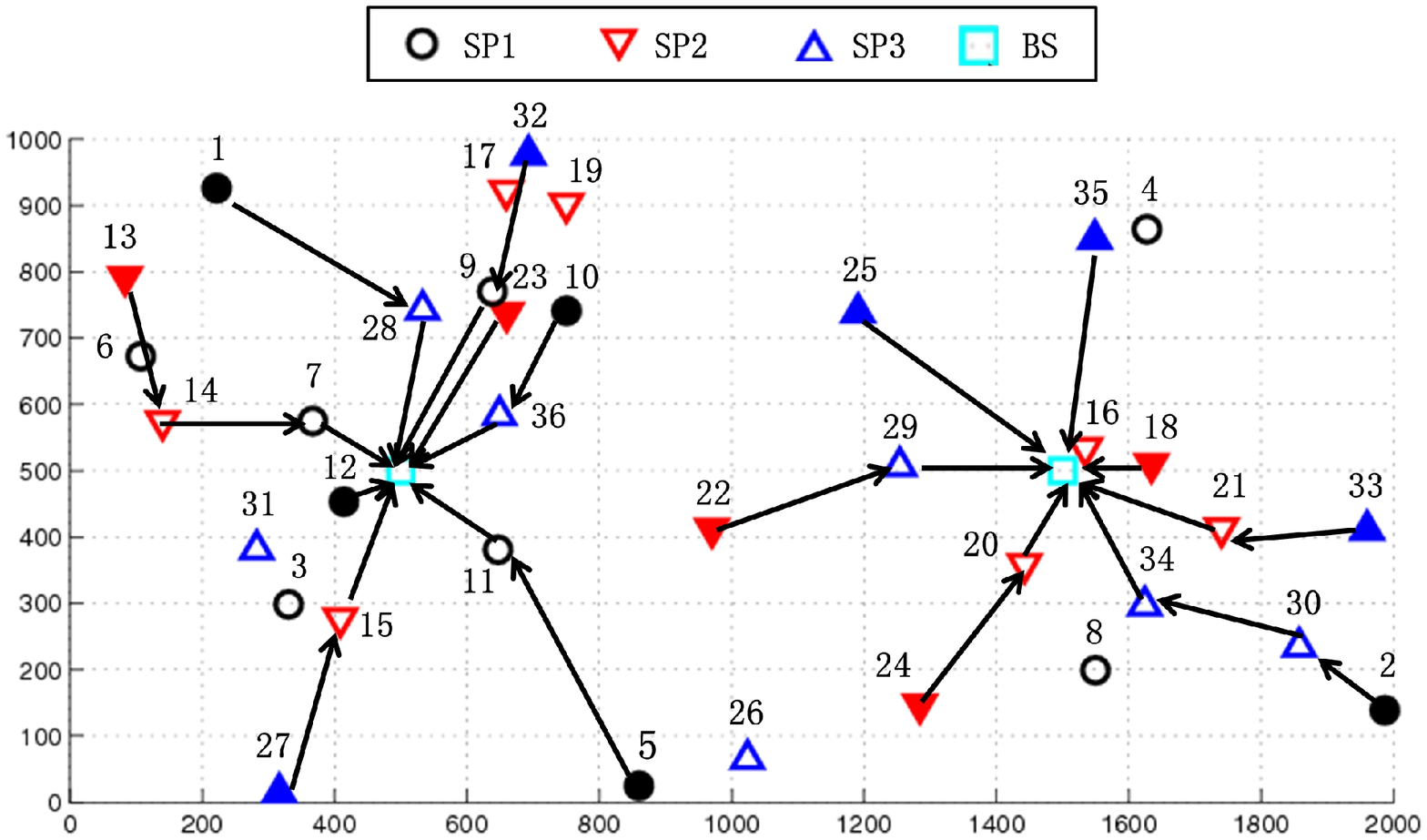}
\caption{Network structure of the WRN (15 Source TDs )} \label{15sources}
\end{figure}

We examine the layered coalitional game in the WRN with uniform distribution of TDs.
In Fig.~\ref{12sources}, $N=36$ TDs are randomly deployed over the entire area of the WRN, with $\mathcal{N}^{SP1}=\{1,2,\ldots,12\}$, $\mathcal{N}^{SP2}=\{13,14,\ldots,24\}$ and $\mathcal{N}^{SP3}=\{25,26,\ldots,36\}$. Each SP first randomly chooses $4$ of its TDs to be the source TDs. The network is also initialized as a star topology in each cell.  The arrowlines in Fig.~\ref{12sources} shows the convergence toward the final Nash forest structure (i.e., a Nash network) resulting from the proposed layered coalitional game. The dot, downtriangle and uptriangle represent the TDs from $SP1$, $SP2$ and $SP3$, respectively, with the filled ones representing the source TDs and empty ones representing vacant TDs. Fig.~\ref{12sources} clearly shows that, through distributed decisions, the source TDs form the multi-hop structure by utilizing the vacant TDs for relay. As we observe from the link connections, the corresponding coalitional structure for the SPs is the grand coalition, i.e., $\{(SP1,SP2,SP3)\}$. Once all the SPs reach the stable coalitional structure, they will remain in its coalition until the environment changes. Then, to show that the proposed algorithm is able to adapt to the environment change (e.g., traffic demand change), we vary the traffic demand by randomly turning a vacant TD from each SP into a source TD. Fig.~\ref{15sources} shows the updated network structure, from which the randomly chosen TDs are TD$10$, TD$22$ and TD$27$. Taking a close look at the new network structure, it can be found that TD$36$ switches from TD$23$ to TD$10$. This is because TD$36$ can help to improve the throughput of TD$10$ by $0.175$, thus $21$ in payoff. However, TD$36$ can only help to increase the throughput of TD$23$ by $0.116$, thus $14$ in payoff. Based on the myopic principle, TD$36$ disconnects from TD$23$ and connects to TD$10$ to improve the total payoff of the WRN. Since TD$23$ has no other option for relaying, it has to perform the direct transmission to the BS. Similarly, TD$29$ disconnects from TD$25$ and connects with TD$22$. It is interesting to observe that, although TD$22$ locates in the left side of the area, it chooses the right BS for the uplink transmission. As the TDs are throughput-oriented, the proposed algorithm can help the TD to find the optimal BS to maximize the end-to-end throughput. Moreover, the proposed algorithm enables TD$27$ to find the optimal relay, i.e., TD$15$, among several possible relay options, e.g., TD$3$, TD$15$ and TD$31$, to maximize the throughput of TD$27$. In the proposed algorithm, it only takes one iteration for TDs to form the new network structure adapting to the traffic demand change. We can observe that the corresponding coalitional structure remains intact, i.e., the grand coalition $\omega_{8}$.

Table~\ref{cc5} shows the results from the utility allocation solution under different coalitional structures.
As expected, the grand coalition, i.e., $\omega_{8}$, resulting from the layered coalitional game is stable. Moreover, as we can observe, the grand coalition is the only stable coalition structure. The grand coalition is stable because none of the three SPs can improve its utility by breaking the cooperation with any others. $\omega_{1}$ is not stable because any two of the SPs can cooperate to gain higher individual utility. $\omega_{2}$ and $\omega_{3}$ are not stable since by transiting to $\omega_{7}$ any one of the SPs is able to gain higher utility. $\omega_{4}$ is not stable because $SP2$ can establish coalition with either $SP1$ or $SP3$ to form $\omega_{5}$ or $\omega_{6}$, respectively, to improve its utility. $\omega_{5}$, $\omega_{6}$ and $\omega_{7}$ are not stable because the two non-cooperative SPs in these coalitional structures can both gain higher individual utility by cooperation, transiting the coalitional structure to $\omega_{8}$.

\begin{figure}
\centering
\includegraphics[width=0.8\textwidth]{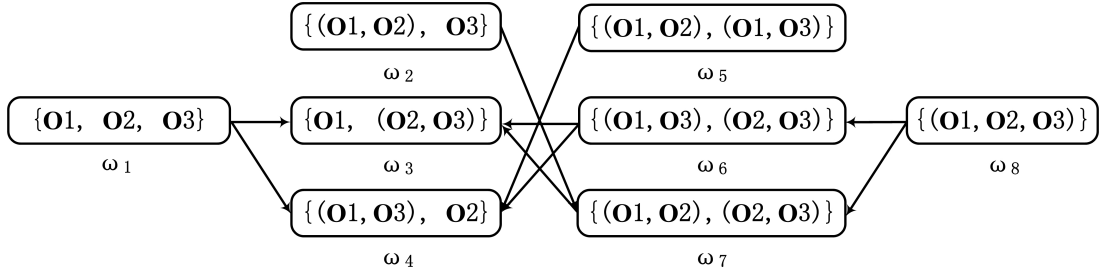}
\caption{State transition diagram of coalitional structures in WRN ($C^{co}=15$).} \label{cost15}
\end{figure}

When the coalition cost is varied, the current coalitional structure may not always be stable and yield the maximum aggregated utility. SPs are free to form any stable coalitional structures that are beneficial to them.  Specifically, we look into the details of cases with coalition cost $C^{co}=15$ and $C^{co}=35$ as shown in Tables~\ref{cc15} and~\ref{cc35}, respectively. In the case with $C^{co}=15$, there are two stable coalitional structures, i.e., $\omega_{3}$ and $\omega_{4}$, under the utility allocation solution. They are stable because none of the SPs can unilaterally improve its utility by splitting from its current coalition(s) or joining new coalition(s). The final stable coalitional structure depends on the initial coalitional structure and the sequence of the coalition formation among the SPs. Taking the case of $C^{co}=15$ for example, we illustrate the transition diagram of three SPs in Fig.~\ref{cost15}. If the initial state is $\omega_{5}$, then $\omega_{4}$ is the only coalitional structure that can be reached when the merge-and-split updating rule is applied. However, if the initial state is $\omega_{6}$, either $\omega_{3}$ or $\omega_{4}$ could be reached depending on which SP, either $SP1$ or $SP2$, will break cooperation with $SP3$ first. Similarly, the final coalitional structure resulting from other initial coalitional structures can be found following the same analysis. In the case of $C^{co}=35$, we can observe that there is only one stable coalitional structure, i.e., $\omega_{1}$. As none of SPs can benefit from forming any coalitional structure due to the high coalition cost, each SP tends to operate without cooperation. With the above analysis, we can observe that the utility allocation solution can help each SP to determine the fair utility share under different possible coalitional structures. With these allocation results, the rational SPs can make decisions to form stable coalitional structures for the purpose of utility improvement.

\section{Conclusion}
The layered cooperation in the wireless relay network (WRN) leads to the improved network capacity for the terminal devices (TDs) and higher benefit for the service providers (SPs). In this paper, we have presented a game theoretic framework for SP and TD cooperation. Specifically, we have proposed a layered coalitional game to model the cooperation behavior among the SPs and TDs in the WRN. The overlapping coalition formation game has been proposed for SPs to find the stable coalitional structures, while the network structure formation game has been introduced for TDs to form the stable network structures. We have shown the convergence of the algorithm at both layers. This paper has also addressed the utility allocation problem of cooperative SPs in the WRN by adopting the concepts of Shapley value. Based on these concepts, we have obtained the utility allocation solution to stabilize the coalitional structure generated by the proposed layered coalitional game under a general network topology.

\newpage

\begin{table*}
\centering
\caption{\footnotesize Utility Matrix for Three-SP Coalitional Game With Coalition Cost $C^{co}=5$. } \label{cc5}
\begin{tabular}{|c|c|c|c|c|}
\hline
Coalitional Structure & $\phi_{1}(\omega) $ & $\phi_{2}(\omega)$ & $\phi_{3}(\omega)$ & $\phi_(\omega)$ \\ \hline
\hline
$\omega_{1}=\{SP1,SP2,SP3\}$ & $390$ & $452$ & $424$ & $1266$ \\
$\omega_{2}=\{(SP1,SP2),SP3\}$ & $407.5$ & $469.5$ & $424$ & $1301$ \\
$\omega_{3}=\{SP1,(SP2,SP3)\}$ & $390$ & $480.5$ & $452.5$ & $1323$ \\
$\omega_{4}=\{(SP1,SP3),SP2\}$ & $402$ & $452$ & $436$ & $1290$ \\
$\omega_{5}=\{(SP1,SP2),(SP1,SP3)\}$ & $403$ & $459$ & $436$ & $1298$ \\
$\omega_{6}=\{(SP1,SP3),(SP2,SP3)\}$ & $408$ & $493$ & $437$ & $1338$ \\
$\omega_{7}=\{(SP1,SP2),(SP2,SP3)\}$ & $416$ & $484$ & $474$ & $1374$ \\
$\omega_{8}=\{(SP1,SP2,SP3)\}$ & $419.5$ & $498$ & $474.5$ & $1392$ \\
\hline
\end{tabular}
\end{table*}

\begin{table*}
\centering
\caption{\footnotesize Utility Matrix for Three-SP Coalitional Game With Coalition Cost $C^{co}=15$. } \label{cc15}
\begin{tabular}{|c|c|c|c|c|}
\hline
Coalitional Structure & $\phi_{1}(\omega) $ & $\phi_{2}(\omega)$ & $\phi_{3}(\omega)$ & $\phi_(\omega)$ \\ \hline
\hline
$\omega_{1}=\{SP1,SP2,SP3\}$ & $390$ & $452$ & $424$ & $1266$ \\
$\omega_{2}=\{(SP1,SP2),SP3\}$ & $397.5$ & $459.5$ & $424$ & $1281$ \\
$\omega_{3}=\{SP1,(SP2,SP3)\}$ & $390$ & $470.5$ & $442.5$ & $1303$ \\
$\omega_{4}=\{(SP1,SP3),SP2\}$ & $392$ & $452$ & $426$ & $1270$ \\
$\omega_{5}=\{(SP1,SP2),(SP1,SP3)\}$ & $383$ & $449$ & $426$ & $1258$ \\
$\omega_{6}=\{(SP1,SP3),(SP2,SP3)\}$ & $398$ & $483$ & $417$ & $1298$ \\
$\omega_{7}=\{(SP1,SP2),(SP2,SP3)\}$ & $406$ & $464$ & $464$ & $1334$ \\
$\omega_{8}=\{(SP1,SP2,SP3)\}$ & $399.5$ & $478$ & $454.5$ & $1332$ \\
\hline
\end{tabular}
\end{table*}

\begin{table*}
\centering
\caption{\footnotesize Utility Matrix for Three-SP Coalitional Game With Coalition Cost $C^{co}=35$. } \label{cc35}
\begin{tabular}{|c|c|c|c|c|}
\hline
Coalitional Structure & $\phi_{1}(\omega) $ & $\phi_{2}(\omega)$ & $\phi_{3}(\omega)$ & $\phi_(\omega)$ \\ \hline
\hline
$\omega_{1}=\{SP1,SP2,SP3\}$ & $390$ & $452$ & $424$ & $1266$ \\
$\omega_{2}=\{(SP1,SP2),SP3\}$ & $377.5$ & $439.5$ & $424$ & $1241$ \\
$\omega_{3}=\{SP1,(SP2,SP3)\}$ & $390$ & $450.5$ & $422.5$ & $1263$ \\
$\omega_{4}=\{(SP1,SP3),SP2\}$ & $372$ & $452$ & $406$ & $1230$ \\
$\omega_{5}=\{(SP1,SP2),(SP1,SP3)\}$ & $343$ & $429$ & $406$ & $1178$ \\
$\omega_{6}=\{(SP1,SP3),(SP2,SP3)\}$ & $378$ & $463$ & $377$ & $1218$ \\
$\omega_{7}=\{(SP1,SP2),(SP2,SP3)\}$ & $386$ & $424$ & $444$ & $1254$ \\
$\omega_{8}=\{(SP1,SP2,SP3)\}$ & $359.5$ & $438$ & $414.5$ & $1212$ \\
\hline
\end{tabular}
\end{table*}

\end{document}